# Studies of scintillation properties of CaMoO$_4$ at millikelvin temperatures

X. Zhang[1], J. Lin[1], V. B. Mikhailik[2,a)] and H. Kraus[1]

[1]*Department of Physics, University of Oxford, Keble Rd., Oxford, OX1 3RH, UK*

[2]*Diamond Light Source, Harwell Science Campus, Didcot, OX11 0DE, UK*

Application of CaMoO$_4$ as a scintillation target in cryogenic rare event searches relies on the understanding of scintillation properties of the material at the temperatures at which these detectors operate. We devised and implemented a detection module with a low-temperature photomultiplier from Hamamatsu (model R8520-06) powered by a Cockcroft-Walton generator. The detector module containing the CaMoO$_4$ crystal was placed in a $^3$He/$^4$He dilution refrigerator and used to measure scintillation characteristics of CaMoO$_4$ in the millikelvin temperature range. At the lowest temperature achieved, the energy resolution of CaMoO$_4$ for 122 keV γ from a $^{57}$Co source is found to be 30% (FWHM), and the fast and slow decay constants are 40.6±0.8 μs and 3410±50 μs, respectively. The temperature variation of the CaMoO$_4$ decay kinetics is discussed in terms of a three-level model of the emission center.

**I. INTRODUCTION**

Interest in scintillating CaMoO$_4$ has increased recently, driven by the prospect of application in cryogenic experiments searching for rare events [1] such as neutrinoless double beta decay (0υ2β) or dark matter particle detection. Some of these experiments are based on a detector technology that involves simultaneous readout of a phonon and a light response induced by particle interaction in a scintillating target [1], [2]. For this technique the excellent energy resolution and low-energy threshold of a phonon detector is combined with event discrimination ability that allows rejection of events caused by background radioactivity [3].

---
[a] Author to whom correspondence should be addressed. Electronic mail: vmikhai@hotmail.com

Recent results obtained by experiments searching for rare events such as interactions with Weakly Interactive Massive Particle (WIMP) dark matter [4], radioactive decay of long-lived nuclei [5] and neutrinoless double beta decay [6], demonstrated the benefits of this detection technique. It should be emphasized that calcium molybdate has particular merit for cryogenic rare event searches; it contains $^{100}$Mo that is considered to be a favourable nucleus for the detection of neutrinoless double beta decay due to its natural abundance and high Q-value. This is one of the reasons for the choice of $CaMoO_4$ for the large scale experiment AMoRE, [7]. Thus, there is significant interest in research into scintillating $CaMoO_4$ and its material properties over a wide range of temperatures [8], [9], [10]

With $CaMoO_4$ being an interesting candidate for a target material in a cryogenic phonon-scintillation detector for rare event searches, scintillation properties have been measured down to 7 K [11], but no systematic studies at temperatures much lower than this have been carried out so far. It is anticipated that the scintillation light yield should not change with further cooling, similar to what has been found for $CaWO_4$ [12]. However, this is not necessarily the case for the decay kinetics. Measurements involving optical excitation demonstrated that in molybdates the luminescence decay rate exhibits more than an order of magnitude increase before it reaches a constant value below 1 K. This is explained by a splitting of the emission level into a pair of closely spaced sublevels [13], [14], [15], [16]. This model can be adopted to explain results on the temperature dependence of the scintillation kinetics in $CaMoO_4$ but the lack of data below 1 K prevented reliable determination of the low-temperature decay rates from the excited levels and their energy splitting. Thus, the objective of extending the characterization of $CaMoO_4$ scintillation to the millikelvin range is twofold: the determination of scintillation efficiency and decay constant at typical operating temperatures of cryogenic phonon-scintillation detectors, and finding details of the energy structure of the emission centre in the crystal.

**II. EXPERIMENT**

Detecting photons from a scintillating crystal, cooled to millikelvin temperature, presents a significant experimental challenge. There are two main approaches to this problem. In most cases, photon sensors, operating at ambient temperature are used, with light from the cooled scintillator sample guided outside the cryostat using windows, light pipes or optical fibres. The drawbacks of this approach are very low light collection efficiency and heating of the sample by blackbody radiation entering the cryostat through the optical passage, and that limits the lowest temperature that can be reached. The second, technically more challenging, approach relies on a photodetector in close proximity to the scintillation crystal inside of a cryostat. This can improve light collection significantly but crucially depends on the photo detector's ability to operate at such low temperatures.

In the experiment, reported here, we implemented the second approach: both low-temperature photomultiplier tube (PMT) and $CaMoO_4$ crystal were placed inside the cryostat and cooled during measurements. This approach can offer advanced opportunities for cryogenic experiments involving scintillators. Excellent sensitivity for single photon detection, large sensitive area, low noise and ruggedness remain compelling reasons for preferring PMTs for the detection of scintillation. Major research efforts have been carried out in the last decade aiming to improve the performance of PMTs and widening the temperature range of their operation. The suitability of a modified (low-temperature) version of PMTs for operation at temperatures of liquid noble gases has already been proven in numerous studies [17] and recently it has been demonstrated that the low-temperature version of such a PMT can be used for detection of single photons when cooled close to the temperature of liquid He [18].

We used in this study the low-temperature photomultiplier model R8520-06 from Hamamatsu, developed for detection of scintillation of liquid xenon [19]. Measurements reported here were carried out in a $^3$He/$^4$He dilution refrigerator model K-400 (Oxford Instruments). The PMT and scintillating crystal were installed in a copper detector housing attached to the bottom of the dilution refrigerator's mixing chamber. The walls of the detector housing were covered with reflective foil VM2000 from 3M. The scintillating $CaMoO_4$ crystal of rectangular shape $5\times10\times10$ mm$^3$ with polished surfaces and a phonon sensor (NTD-Ge) glued to one side was suspended within the housing using nylon fibres. The voltages for the dynode chain for the PMT were produced using a Cockcroft-Walton generator [20]. Scintillation signal detected by PMT is digitized by ADC with 5 ns sampling interval that allows resolving individual photons. The measurements and analysis of scintillation events were carried out using the multi-photon counting technique described elsewhere [21].

## II. RESULTS AND DISCUSSION

In the experiment reported here, the $CaMoO_4$ sample was cooled to temperatures as low as 17 mK, with temperatures below 100 mK deduced from the resistance of the NTD-Ge sensor attached to the sample, using a generic calibration [22]. At higher temperature, this resistance becomes too small to give sensible readings, while a ruthenium oxide thermometer integrated within the K400 cryostat is within its reliable range. Hence the temperatures above 100 mK were obtained from the K400 operations monitoring system, which uses a generic, calibrated 2200 Ω $RuO_2$ resistor to measure the mixing chamber temperature. Figure 1 shows the histogram of the number of photons per scintillation event detected in $CaMoO_4$ when excited by γ-quanta, emitted from a $^{57}Co$ radioactive source placed outside of the cryostat. The histogram exhibits a clear peak which corresponds to a mean number of 60 scintillation photons, detected by the PMT, following interaction of 122 keV γs with the scintillator crystal.

The ratio of the full width at half maximum (FWHM) to the peak position, which is a measure of the energy resolution of the scintillation detector, is assessed to be (18/60)×100%=30%. Although the resolution is dominated by the contribution of the photon statistics and only average as far as scintillation detectors are concerned, it is nevertheless similar to the energy resolution of a $B_4G_3O_{12}$ scintillator at room temperature [23].

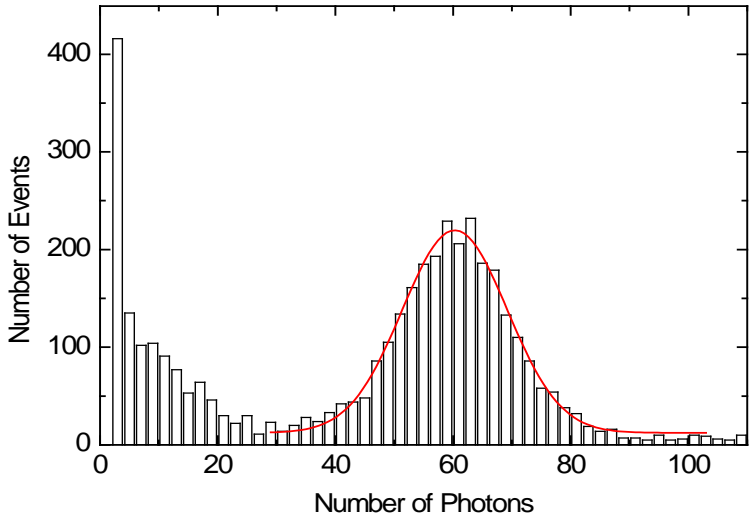

Fig. 1. Histogram of the number of photons per scintillation event detected in $CaMoO_4$ at T=17 mK under excitation with γ from a $^{57}Co$ source. The solid line represents a fit of a Gaussian to the peak.

Previously we investigated scintillation properties of $CaMoO_4$ down to 7 K only, due to technical limitations of the He-flow cryostat used [11]. The present study moved this low temperature limit by almost three orders of magnitude: the decay curve of $CaMoO_4$ shown in Fig. 2 was obtained from the analysis of the time structure of scintillation events recorded at 17 mK. A fit to the experimental data was carried out using a sum of two exponential functions.

The fast and slow decay constants at this temperature are found to be $\tau_f = 40.6 \pm 0.8$ μs and $\tau_s = 3410 \pm 50$ μs. To investigate temperature dependence, we measured scintillation decay curves at different temperatures of the $CaMoO_4$ crystal up to 6.4 K. Thus, when combined with the results of our earlier measurements of this crystal in the 7 – 295 K temperature range [11] these new data allowed investigation of $\tau = f(T)$ for the calcium molybdate scintillator (see Fig. 3) over a much wider temperature range. At low temperatures, both decay constants $\tau_f$ and $\tau_s$ are constant but they start to decrease rapidly at around 1 K and over a 10 K temperature range they reduce by about one order of magnitude. Beyond that, the scintillation kinetic experiences little change over a wide temperature range. This changes abruptly above 180 K when the slow decay time constant shows a rapid decrease due to processes of thermally activated quenching.

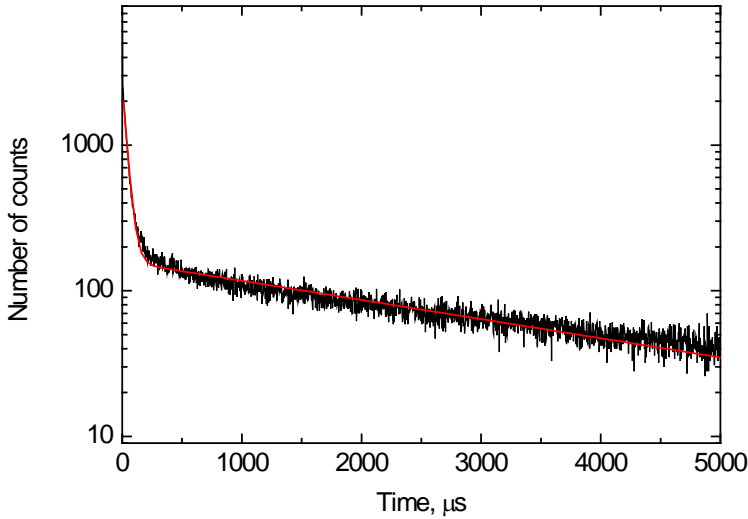

Fig. 2. Scintillation decay curve of $CaMoO_4$ measured at T = 17 mK for γ-excitation ($^{57}Co$). The solid curve shows the best fit to the experimental data using the sum of two exponential functions.

The observed temperature dependence can be readily explained in the framework of the model of the emission centre constituting a ground state and two excited levels of which the lower one is a metastable level. The basics of this model have been initially developed for the interpretation of features of impurity emission in solids [24], [25] and lately this has been used to explain decay kinetics of molybdates and tungstates [12], [13], [14], [15], [16]. Here we applied this model to the interpretation of the features observed in the $\tau = f(T)$ dependency of the $CaMoO_4$ scintillator.

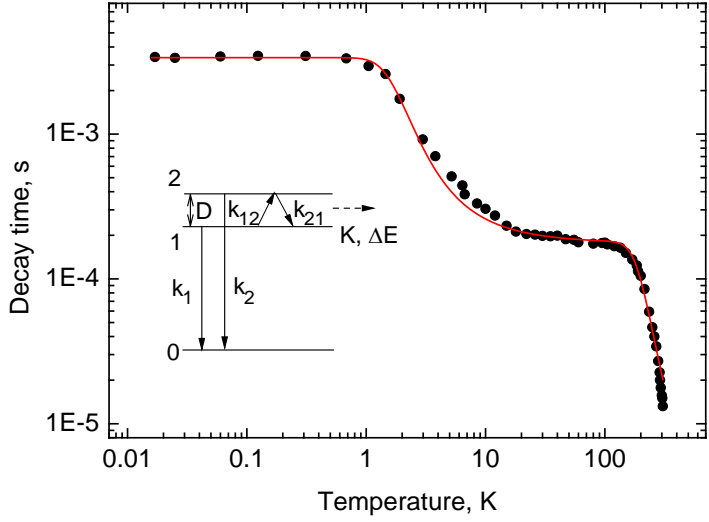

Fig. 3. Temperature dependence of the slow scintillation decay time constant of $CaMoO_4$. The solid curve displays the best fit to the experimental data using the three-level model shown in the insert. The parameters of the model are as follows: $k_1$ and $k_2$ - radiative rates of the lower and upper states respectively, $k_{12}$ and $k_{21}$ are non-radiative rates between these levels, $D$ - energy gap between two levels, $K$ - frequency factor and $\Delta E$ - activation barrier energy.

The luminescence of $CaMoO_4$ is due to the radiative recombination of self-trapped excitons (STE), localised at the oxyanion complex $MoO_4^{2-}$. At low temperatures it exhibits a broad band with a maximum at around 540 nm [26], [27]. The emission transitions originate from the two triplet levels $^3T_2$ and $^3T_1$ and terminate at the $^1A_1$ ground state. Thus, both triplet levels contribute to the emission that results in the spectrum and decay kinetics composed of few components [8], [11], [27] [28]. The energy level scheme of the emitting centre is sketched in the insert of Fig.3. The probabilities of radiative transitions from levels 1 and 2 are denoted as $k_1$ and $k_2$ respectively. The emitting levels are separated by the energy gap $D$ and it is assumed that non-radioactive transitions between these levels $k_{12}$ and $k_{21}$ are single phonon processes [29]. The model also accounts for the possibility of the emission centre to decay non-radiatively through a thermally activated process with a rate $k_x = K\exp(-\Delta E/kT)$, where $K$ is the frequency factor and $\Delta E$ is the activation barrier. The expression for the temperature dependence of slow decay time constant ($\tau_s$) was derived in [12] as following:

$$\frac{1}{\tau_s} = \frac{k_1 + k_2 \exp(-D/kT)}{1 + \exp(-D/kT)} + K\exp(-\Delta E/kT) \quad (1)$$

This formula was used to fit the experimental results as shown in Fig. 3. It should be noted that the fitting procedure is very sensitive to the selection of initial parameters, the convergence is slow and the values of some parameters are obtained with significant errors (for example the error on K is especially large, $14\times10^6$). The energy parameters are derived, though, with fairly small errors, permitting precise reconstruction of the energy structure of the emission centre. The obtained values are summarised in Table I, together with published parameters obtained from the fit of $\tau = f(T)$ dependences for other molybdates and tungstates.

Table I. Parameters of emission centre in CaMoO$_4$ (and for comparison, other molybdates and tungstates) obtained from fitting the temperature dependence of decay time constants: $k_1$ and $k_2$ - radiative rates of the lower and upper states respectively, $D$ - energy gap between two levels, $K$ - frequency factor and $\Delta E$ - activation barrier energy.

| Crystal | $k_1$, s$^{-1}$ | $k_2$, s$^{-1}$ | $D$, meV | $K$, s$^{-1}$ | $\Delta E$, meV | Reference |
|---|---|---|---|---|---|---|
| CaMoO$_4$ | 298±2 | 11400±700 | 0.61±0.01 | $6.1\times10^6$ | 126±47 | this work |
| CaMoO$_4$ | 330 | 8000 | 0.6±0.1 | - | - | [13]* |
| CdMoO$_4$ | 40 | 8000 | 0.07 | - | - | [15*] |
| PbMoO$_4$ | 6300 | 130000 | 0.37 | $1\times10^5$ | 260 | [14]* |
| CdWO$_4$ | 3120 | 95000 | 1.2 | $1.2\times10^5$ | - | [14]* |
| ZnWO$_4$ | 4980 | 58000 | 1.0 | $1\times10^5$ | 290 | [14]* |
| ZnWO$_4$ | 5000 | 33300 | 2.1 | $2\times10^9$ | 300 | [16]* |

*- UV excitation.

The energy gap between the two emitting levels $D$ in molybdates is smaller than the respective value in tungstates. This may be indicative of a smaller crystal field effect that causes the splitting of energy levels in the crystals [28].

**VI. CONCLUSION**

Potential applications of calcium molybdate crystals in detectors for low-temperature cryogenic experiments searching for rare events require information on scintillation properties of this material at millikelvin temperatures. The achievable energy threshold and resolution of such detectors relies on the light yield of the crystal at this temperature while the scintillation decay constant affects the detector dead time and coincidence window. These parameters are important for evaluating the sensitivities of experiments that use $CaMoO_4$ cryogenic scintillators. Currently these evaluations are based mostly on the data obtained at around 10 K and on the assumption that scintillation properties do not change with further cooling.

Motivated by this we developed the experimental setup for measuring scintillation characteristics of crystals in the millikelvin temperature range. It uses a low-temperature PMT from Hamamatsu (model R8520-06) powered by a Cockcroft-Walton generator. By means of this equipment we detected scintillations from a $CaMoO_4$ crystal cooled to 17 mK. This allowed characterisation of the performance of the scintillator in the range of temperatures where cryogenic phonon-scintillation detectors operate. The decay kinetics changes drastically upon cooling to below 10 K when the decay constant exhibits an increase by an order of magnitude. We analysed the temperature dependence of $\tau = f(T)$ using a three-level model to describe the emission center and derived quantitative parameters for the model.

This study also demonstrated that low-temperature PMT can be used for detection of scintillation at millikelvin temperatures.

**ACKNOWLEGMENT**

This study was partially supported by a grant from Science & Technology Facilities Council (STFC) and by a grant from the Royal Society (London) ''Cryogenic scintillating bolometers for priority experiments in particle physics''.

We would also like to thank Dr Xavier-Francois Navick, CEA Saclay, for lending us the NTD-Ge sensor used in experiment.